\documentclass[prd,twocolumn,showpacs,preprintnumbers,amsmath,amssymb]{revtex4}
\usepackage{graphicx}% Include figure files
\usepackage{dcolumn}% Align table columns on decimal point
\usepackage{bm}% bold math
\usepackage{psfig}
\usepackage{epsfig}

\begin{document}

\newcommand{\vAi}{{\cal A}_{i_1\cdots i_n}}
\newcommand{\vAim}{{\cal A}_{i_1\cdots i_{n-1}}}
\newcommand{\vAbi}{\bar{\cal A}^{i_1\cdots i_n}}
\newcommand{\vAbim}{\bar{\cal A}^{i_1\cdots i_{n-1}}}
\newcommand{\htS}{\hat{S}}
\newcommand{\htR}{\hat{R}}
\newcommand{\htI}{\hat{I}}
\newcommand{\htB}{\hat{B}}
\newcommand{\htD}{\hat{D}}
\newcommand{\htV}{\hat{V}}
\newcommand{\cT}{{\cal T}}
\newcommand{\cM}{{\cal M}}
\newcommand{\cMs}{{\cal M}^*}
\newcommand{\vk}{{\bf k}}
\newcommand{\vK}{{\\bf K}}
\newcommand{\vb}{{\textstyle{\bf b}}}
\newcommand{{\vp}}{{\vec p}}
\newcommand{{\vq}}{{\vec q}}
\newcommand{\vQ}{{\vec Q}}
\newcommand{\vx}{{\textstyle{\bf x}}}
\newcommand{\vbb}{{\textstyle{\bf b}}}
\newcommand{\vv}[1]{{\textstyle{\bf v_{#1}}}}
\newcommand{\vj}{{\textstyle{\bf j}}}
\newcommand{\vE}{{\textstyle{\bf E}}}
\newcommand{\invvol}{{\frac{d\omega d\vk}{(2\pi)^{4}}}}
\newcommand{\tr}{{{\rm Tr}}}
\newcommand{\beq}{\begin{equation}}
\newcommand{\eeq}[1]{\label{#1} \end{equation}}
\newcommand{\half}{{\textstyle \frac{1}{2} }}
\newcommand{\lton}{\mathrel{\lower.9ex \hbox{$\stackrel{\displaystyle
<}{\sim}$}}}
\newcommand{\gton}{\mathrel{\lower.9ex \hbox{$\stackrel{\displaystyle
>}{\sim}$}}}
\newcommand{\ee}{\end{equation}}
\newcommand{\ben}{\begin{enumerate}}
\newcommand{\een}{\end{enumerate}}
\newcommand{\bit}{\begin{itemize}}
\newcommand{\eit}{\end{itemize}}
\newcommand{\bc}{\begin{center}}
\newcommand{\ec}{\end{center}}
\newcommand{\bea}{\begin{eqnarray}}
\newcommand{\eea}{\end{eqnarray}}
\newcommand{\beqar}{\begin{eqnarray}}
\newcommand{\eeqar}[1]{\label{#1}\end{eqnarray}}
\newcommand{\bra}[1]{\langle {#1}|}
\newcommand{\ket}[1]{|{#1}\rangle}
\newcommand{\norm}[2]{\langle{#1}|{#2}\rangle}
\newcommand{\brac}[3]{\langle{#1}|{#2}|{#3}\rangle}
\newcommand{\hilb}{{\cal H}}
\newcommand{\pleft}{\stackrel{\leftarrow}{\partial}}
\newcommand{\pright}{\stackrel{\rightarrow}{\partial}}

\begin{flushright}
%DRAFT 4/7 10pm
%\vskip .5cm
\end{flushright} \vspace{1cm}

\title{Collisional Energy Loss of Non Asymptotic Jets in a QGP}

\author{A.~Adil}%
\email{azfar@phys.columbia.edu}

\author{M.~Gyulassy}
\email{gyulassy@phys.columbia.edu}

\author{W.~Horowitz}
\email{horowitz@phys.columbia.edu}

\author{S.~Wicks}
\email{simonw@phys.columbia.edu}

\affiliation{Columbia University, Department of
Physics, 538 West 120-th Street, New York, NY 10027
}%

\date{\today}% It is always \today, today,

\begin{abstract}
We calculate the collisional energy loss suffered by a heavy (charm) quark created at a finite time within a Quark Gluon Plasma (QGP) in the classical linear response formalism as in Peigne {\it et al.} \cite{peigne}.  We pay close attention to the problem of formulating a suitable current and the isolation of binding and radiative energy loss effects.  We find that unrealistic large binding effects arising in previous formulations must be subtracted.  The finite time correction is shown to be important only for very short length scales on the order of a Debye length.  The overall energy loss is similar in magnitude to the energy loss suffered by a charge created in the asymptotic past.  This result has significant implications for the relative contribution to energy loss from collisional and radiative sources and has important ramifications for the ``single electron puzzle'' at RHIC. 
\end{abstract}

\pacs{12.38.Mh; 24.85.+p; 25.75.-q}

\maketitle

%%%%%%%%%%%%%%%%%%%%%%%%%%%%%%%%%%%%%%%%%%%%%%%%%%%%%%%%%%%%%%%%%%%%%%%%%%
\section{Introduction}

One of the experimental signatures of the creation of novel QCD matter at the Relativistic Heavy Ion Collider (RHIC) is the detailed suppression pattern of high transverse momentum hadrons \cite{JetQuench}.  This ``jet quenching'' has been previously explained as caused by the energy loss due to induced gluon bremmstrahlung in a hot Quark Gluon Plasma (QGP) \cite{Gyulassy:2003mc}.  Recent data for the suppression of single non photonic electrons \cite{singelec} (thought to probe the energy loss of heavy quarks), however, cannot be reproduced via purely radiative energy loss arguments with a physically reasonable set of parameters \cite{djordjwicks}.

A recent publication by Mustafa \cite{mustafa} stated that a reasonable set of parameters at RHIC would lead to elastic energy loss of a heavy quark that is of the same order as radiative energy loss.  This conclusion goes against previous estimates of the collisional energy loss which used asymptotic arguments to assert that the radiative energy loss is much larger than elastic \cite{bjorken}.  Inspired by this development, a recent paper by Wicks {\it et al.} \cite{wicks}  gets better agreement with the single non photonic electron data by incoherently adding the energy loss due to elastic as well as radiative sources while also averaging over geometric and gluon number fluctuations.  It is extremely important to take fluctuations into account as they allow for the simultaneous reproduction of both the light quark suppression and the non photonic electron data.  Their paper, however, neglects the finite time creation effects on the collisional part of the energy loss that arise from the fact that the heavy quark probe is part of a back to back hard scattering occurring at a finite time in the medium.

Peigne {\it et al.} \cite{peigne} were the first to study the effects of the finite creation time of the probe on the collisional energy loss suffered by it.  They claimed that the collisional energy loss was largely suppressed in comparison to the infinite limit for medium lengths applicable to RHIC physics.  In the current paper, we redo the collisional energy loss calculation in the same formulation as Peigne {\it et al.}, paying close attention to the exclusion of all radiative effects and especially binding effects that arise if the full color current is approximated incorrectly.

Recent papers by Djordjevic \cite{djordj} and Wang \cite{Wang} address this problem using a quantum mechanical formalism rather than the linear response approach used in the current paper.  The effect is modeled by restricting the interaction Hamiltonian to a finite time range in Djordjevic \cite{djordj} while Wang \cite{Wang} considers the interference between radiative and elastic amplitudes in first order medium scattering.  The results from these two papers are inconsistent, however, since the full medium propagators suppress interference between radiative and elastic amplitudes \cite{djordj}.  Our results are consistent with Djordjevic \cite{djordj}.

In Section \ref{Sec1} we describe the subtle problems with formulating and interpreting currents that are created at a finite time.  In Section \ref{Sec2} we formulate and derive calculations for the infinite time case so as to show agreement with previous calculations and illustrate the subtleties in our specific method.  In Section \ref{Sec3} we reproduce the answer derived by Peigne {\it et. al.} and then correct it for radiative and binding effects.  In Section \ref{Sec4} we present a summary of our results.

\section{A Comment on Conserved Currents} \label{Sec1}

When one calculates elastic energy loss in the formalism of a classical macroscopic field theory, the general method is to posit a source current which travels in the medium in question and induces an electric field.  The induced electric field acts back onto the current and does work on it, leading to a decrease in the energy of the source particle.

In our application to jet production in nuclear collisions involves modeling a $2\rightarrow 2$ collision process in pQCD, with one of the final state particles being detected as a hadron after losing energy in the medium.  A classical current that corresponds to this is,
\begin{eqnarray}
j^{\mu a}_{phys}= \Theta(t)\left( q_1^{a}v_1^\mu \delta(\vx -\vv{1}t )+q_2^a v_2^\mu \delta(\vx - \vv{2}t-\vbb)\right) \nonumber \\
+  \Theta(-t)\left( q_1^{a}v_1^{'\mu} \delta(\vx -\vv{1}'t )+q_2^a v_2^{'\mu} \delta(\vx - \vv{2}'t-\vbb)\right). \label{physcurrent}
\end{eqnarray}
The current in Eq. \ref{physcurrent} has two particles with color charge $q_{1}^a$ and $q_2^a$ (Defined via $q_i^a q_j^a=\delta_{ij}C_{R_{i}}\alpha_s$) that experience a collision at $t=0$ with impact parameter $\vb$.  $C_{R_{i}}$ is the quadratic Casimir for representation $R_{i}$ ($C_{R_{i}}=\frac{4}{3}(3)$ for quarks (gluons)), $\alpha_{s}$ is the QCD coupling constant (set to $\alpha_{s}=0.3$) while $v_{i}^{\mu}=\partial_{0}x^{\mu}_i=(1,\vv{i})$ is the four velocity of the $i^{th}$ particle.  The impact parameter is determined by the momentum transfer in the hard collision $Q$, the `primed' velocities are the pre collision velocities of the particles, the `unprimed' velocities are the post collision velocities of the particles consistent with energy and momentum conservation, and the final color configuration for the particles is the same as the initial configuration due to the Abelian approximation in force throughout this paper.

We proceed with the analysis by calculating the induced field due to Eq.~\ref{physcurrent} and act with it on the particle whose energy loss is in question (say particle 1).  The induced field will include creation radiation and interactions between particle 2 and particle 1 alongside the effect that we hope to calculate: the induced field of particle 1 acting back onto particle 1.  The radiative energy loss correction in QCD was first investigated in \cite{termik} and will be removed from consideration in this paper.  The effect of particle 2 on particle 1, what we call the {\it binding} or {\it shared} interaction, averages out to zero in the following way.  In order to calculate a single particle observable one has to sum over all the color orientations of particle 2 for each orientation of particle 1.  This summation cancels the effects of the interaction and gives a zero effect on average.

Due to the complicated nature of the current in Eq.~\ref{physcurrent}, early calculations \cite{tg} in the induced field formalism ignored the effects of radiation, binding and finite time of the hard interactions and used a simple infinite time single particle current.
\begin{equation}
j^{\mu a}_{\infty,(1)}(\vx,t)=q^{a} v^{\mu}\delta(\vx - \vv{} t)
\label{eq:infcurr}
\end{equation}
%For numerical results in the current paper we will concentrate on heavy quark jets and will set $\alpha_{s}=0.3$.
The subscripts in Eq.~\ref{eq:infcurr} specify that this is an infinite time current that was created in the asymptotic past and that there is only one particle present.  The current can also be written in Fourier space,
\begin{equation}
j^{\mu a}_{\infty,(1)}(\omega,\vk)=2\pi q^{a}v^{\mu}\delta(\omega-\vk\cdot\vv{}).
\end{equation}
One can immediately see that the continuity equation written in Fourier space is satisfied, i.e. $k_{\mu}j^{\mu}_{\infty,(1)}=0$ where $k^{\mu}=(\omega,\vk)$.

In order to reintroduce finite time creation effects into the problem, we write down the naive generalization of Eq.~\ref{eq:infcurr},
\begin{eqnarray}
j^{\mu a}_{(1)}(\vx,t)&=&q^{a} v^{\mu}\delta(\vx - \vv{} t)\Theta(t) \nonumber \\
j^{\mu a}_{(1)}(\omega,\vk)&=&iq^{a}\left( \frac{v^{\mu}}{\omega-\vk\cdot\vv{}+i\eta}\right).
\end{eqnarray}
Note, however, that the one particle finite time current is explicitly not conserved, as $\stackrel{\lim}{_{\eta\rightarrow 0 }}k_{\mu}j^{\mu a}_{(1)}=iq^{a}$.  This is expected because creation of charge out of the vacuum is a violation of charge conservation.  

The simplest generalization to a conserved finite time current is,
\begin{equation}
j^{\mu a}_{(2)}(\vx,t)=q^{a} (v_{1}^{\mu}\delta(\vx - \vv{1} t)-v_{2}^{\mu}\delta(\vx - \vv{2} t))\Theta(t).
\end{equation}
This current models a current neutral pair production at $t=0$.  Two oppositely charged particles are created at $t=0$ and propagate outwards from the origin at possibly distinct velocities $\vv{1}$ and $\vv{2}$.  The Fourier space expression for this model current is,
\begin{equation}
j^{\mu a}_{(2)}(\omega,\vk)=iq^{a}\left( \frac{v_{1}^{\mu}}{\omega-\vk\cdot\vv{1}+i\eta}-\frac{v_{2}^{\mu}}{\omega-\vk\cdot\vv{2}+i\eta}\right)
\label{eq:twopart}
\end{equation}
The two particle finite time current satisfies charge conservation and one can explicitly see that $\stackrel{\lim}{_{\eta\rightarrow 0 }}k_{\mu}j^{\mu a}_{(2)}=0$.  The values of the velocity cancel out in the calculation of the charge conservation condition so we can always set $\vv{1}=\vv{}$ and $\vv{2}=0$.  This changes the current to the following:
\begin{eqnarray}
j^{0 a}_{(2)}(\omega,\vk)&=&iq^{a}\left( \frac{1}{\omega-\vk\cdot\vv{}+i\eta}-\frac{1}{\omega+i\eta}\right)\nonumber \\
&=& \frac{iq^{a}}{\omega+i\eta} \frac{\vk\cdot\vv{}}{\omega-\vk\cdot\vv{}+i\eta},\nonumber \\
\vj^{a}_{(2)}(\omega,\vk)&=&iq^{a}\frac{\vv{}}{\omega-\vk\cdot\vv{}+i\eta}.
\label{eq:finv20}
\end{eqnarray}
Eq.~\ref{eq:finv20} was used as the current by Peigne {\it et al.} \cite{peigne}.  

Note that in formulating a suitable finite time current we were forced by charge conservation to include a second particle along with all the complications of the binding interaction.  In order to study purely the finite time effects, we will now have to remove the unrealistic binding terms as well as radiative effects.  Recall that in the actual physical situation corresponding to Eq.~\ref{physcurrent} the binding effect averages out to 0 because the two particles have uncorrelated charge in scattering of color neutral hadrons.  In Eq.~\ref{eq:finv20} , however, the creation of the second particle out of the vacuum leads to exact anti correlation of its charge with the first particle, giving a large unrealistic binding.  This effect is further enhanced by the fact that the second particle is created as a stationary charge at the origin which means that its fields are not relativistically contracted and its effects are longer range and stronger.  We will show below that the enhanced unrealistic binding to the second particle is the cause of the large suppression effect claimed in Peigne {\it et al.} \cite{peigne}.

We will use Eq.~\ref{eq:finv20} as our current when doing the finite time analysis, taking great care to differentiate between self interactions of particles and shared two particle interactions.  We also define an infinite time two particle current to further illustrate the effect of the interaction between multiple particles while avoiding the subtleties of the finite time problem.  
\begin{eqnarray}
j^{0 a}_{\infty,(2)}(\omega,\vk)&=&2\pi q^{a}\left( \delta(\omega-\vk\cdot\vv{})-\delta(\omega)\right) \nonumber \\
\vj^{a}_{\infty,(2)}(\omega,\vk)&=&2\pi q^{a} \vv{}\delta(\omega-\vk\cdot\vv{})
\label{eq:infv20}
\end{eqnarray}

\section{Collisional Energy Loss with Infinite Time Currents} \label{Sec2}

One can write down the rate of change of energy suffered by a particle traveling in a linear medium due to the induced field as,
\begin{equation}
\frac{dE}{dx}=\int \invvol e^{-i(\omega-k v \mu)\frac{L}{v}}\frac{q^{a}}{v}\vv{}\cdot \vE^{a}_{ind}.(\omega,\vk)
\label{eq:dEdx}
\end{equation}
It is understood that $v=|\vv{}|$, $\mu = \vk \cdot \vv{}/kv$ and $L$ is the path length that the particle has traversed in the medium.  In order to regulate classical divergences (due to the Coulomb logarithm), the integration over Fourier modes will only be carried out to a maximum value $k_{max}$.  We assume $k_{max}$ is the maximum momentum transfer in a $2\rightarrow2$ collision process between the incident particle and a medium parton in the center of mass frame,  
\begin{equation}
k_{max}=\min\left\{ E, \frac{2 q (E+p)}{\sqrt{M^{2}+2 q (E+p)}}\right\}.
\label{eq:kmax}
\end{equation}
$E$ is the initial energy of the incident particle, $M$ is its mass, $p$ is its momentum and $q$ is the typical momentum of a medium particle, all in the medium (plasma) frame.  For numerical calculations in this paper, we will consistently use $M=1.2$ GeV in order to be representative of the charm quark energy loss and set $q\sim T=250$ MeV.  %There is some freedom in setting $q\sim T - 3T$.  We use $q\sim T$ because the end result has only a logarithmic dependence on $k_{\max}$ and variations that are still on the order of $T$ give very  similar results.

The induced electric field is the total medium field minus the vacuum field produced by the current, $\vE^{a}_{ind} =\vE^{a}-\vE^{a}_{vac}$.  We can use the calculations in Appendix \ref{App1} to write the following expression for the produced fields.
\begin{eqnarray}
\vE^{a}&=& D_{R}^{\perp}(\omega,\vk)(1-\hat{\vk}\hat{\vk})\vj^{a}+\hat{\vk} D_{R}^{\|}(\omega,\vk)j^{0a} \nonumber \\
\vE^{a}_{vac}&=& D_{R}(\omega,\vk)(\omega \vj^{a} -\vk j^{0 a})
\label{eq:fields}
\end{eqnarray}
The propagators $D_{R}(\omega,\vk)$, $D_{R}^{\perp}(\omega,\vk)$ and $D_{R}^{\|}(\omega,\vk)$ are the vacuum Maxwell propagator, the in medium transverse propagator and the in medium longitudinal propagator respectively, all with retarded pole prescriptions and defined in Appendix \ref{App1} in terms of the Hard Thermal Loop (HTL) polarization functions, $\Pi_{\|}(\omega,\vk)$ and $\Pi_{\perp}(\omega,\vk)$.

\subsection{Infinite Time One Particle Current}

We first calculate the energy loss in the infinite time one particle case in order to compare with previous calculations \cite{bjorken,bt,tg}.  This means using Eq.~\ref{eq:infcurr} as the source current in Eq.~\ref{eq:fields}.  Instead of getting an explicit expression for $\vE^{a}_{ind}$, we separately calculate the total medium energy loss and vacuum energy loss felt by the particle and take the difference at the end.

The current dependence of the vacuum field can be written as $\omega \vj^{a}_{\infty,(1)} -\vk j^{0 a}_{\infty,(1)}=2\pi q^{a}(\omega\vv{}-\vk)\delta(\omega-\vk\cdot\vv{})$.   Using the expression for $D_{R}(\omega,\vk)$ from Appendix \ref{App1},
\begin{eqnarray}
\frac{dE_{vac}}{dx}&=&i\frac{C_{F}\alpha_{s}}{\pi}(1-v^{2})\int dk d\mu \frac{k\mu}{v^{2}\mu^{2}-1}\nonumber \\
&=&0.
\end{eqnarray}
The angular integration leads to the vacuum energy loss being equal to zero.

We can express the electric field in Eq.~\ref{eq:fields} as the following.
\begin{equation}
\vE^{a}(\omega,\vk)=\left( D_{R}^{\|}(\omega,\vk)\frac{k}{\omega}\hat{\vk}\hat{\vk}+ D_{R}^{\perp}(\omega,\vk)(1-\hat{\vk}\hat{\vk})\right)\vj^{a}_{\infty,(1)}
\end{equation}
We use the expressions in Appendix \ref{App1} for the longitudinal and transverse medium retarded propagators and have used continuity of the current in order to write the $0^{th}$ component of the current in terms of the longitudinal current.  This expression for the field can be entered into the formula for energy loss.
\begin{equation}
\frac{dE}{dx}=-i\frac{C_{F}\alpha_{s}}{v2\pi^{2}}\int d\vk\frac{\vk\cdot\vv{}}{k^{2}}\left(\frac{1}{\epsilon_{\|}} -\frac{k^{2}v^{2}-(\vk\cdot\vv{})^{2}}{k^{2}-(\vk\cdot\vv{})^{2}\epsilon_{\perp}}\right)
\end{equation}
$\epsilon_{\|}$ and $\epsilon_{\perp}$ are the longitudinal and transverse dielectric function respectively and are defined in Appendix \ref{App1}.  Both the dielectric functions are evaluated at $\omega=\vk\cdot\vv{}+ i\eta$.  This expression can be rewritten as the following.
\begin{eqnarray}
\frac{dE}{dx}&=&-\frac{C_{F}\alpha_{s}}{v2\pi^{2}}\int d\vk \frac{\vk\cdot\vv{}}{k^{2}}(h_{\|}(\vk)+h_{\perp}(\vk))\nonumber \\
h_{\|}(\vk)&=&\frac{Im(\epsilon_{\|})}{|\epsilon_{\|}|^{2}}\nonumber \\
h_{\perp}(\vk)&=&\frac{(k^{2}v^{2}-(\vk\cdot\vv{})^{2})(\vk\cdot\vv{})^{2}Im(\epsilon_{\perp})}{|(k^{2}-(\vk\cdot\vv{})^{2}\epsilon_{\perp}|^{2}}
\label{eq:preleadlog}
\end{eqnarray}
The parts of the integral that depend explicitly on the real parts of the dielectric function vanish by symmetry.

\begin{figure}
\centering
 \epsfig{file=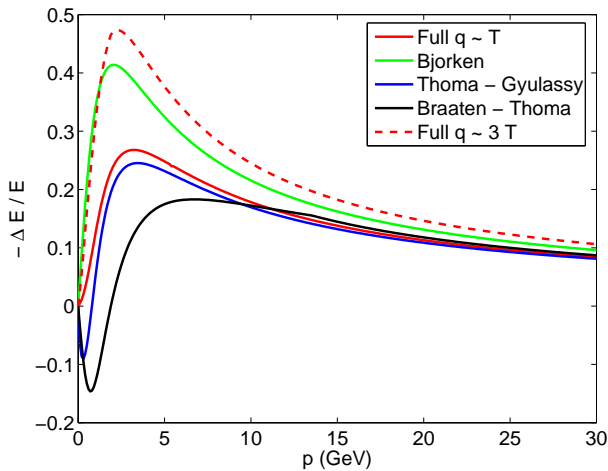,width=3.5in,angle=0}
  \caption{(Color Online) The figure shows the energy loss fraction $-\frac{\Delta E}{E}=- L \frac{dE/dx}{E}$ as a function of the incident momentum $p$.  Curves are shown for an incident charm quark with $M=1.2$ GeV and $L=5$ fm.  It includes the full calculation as well as previous approximations from the literature; Bjorken \cite{bjorken}, Thoma-Gyulassy \cite{tg} and Braaten-Thoma \cite{bt}.}
 \label{fig:infonedeltE}
\end{figure}

In the leading logarithm approximation (as in \cite{tg}) we neglect the effects of the Debye screening mass in the denominator of the integrand in Eq. \ref{eq:preleadlog} and evaluate the integral analytically.\begin{eqnarray}
\frac{dE}{dx}&=&-\frac{C_{F}\alpha_{s}}{2}m_{D}^{2}\log\left(\frac{k_{max}}{m_{D}}\right)f(v)\nonumber \\
f(v)&=&\frac{1}{v^{2}}\left(v-\frac{1}{2}(1-v^{2})\log\left(\frac{1+v}{1-v}\right)\right).
\label{eq:leadlog}
\end{eqnarray}
Here $m_{D}^{2}=4\pi \alpha_{s}T^{2}(1+N_{F}/6)$ is the squared Debye screening mass.  We use $N_{F}=0$ for calculational purposes which is equivalent to assuming that the plasma is purely gluonic.  Notice that the infrared singularity that was regulated by the full dielectric functions in Eq.~\ref{eq:preleadlog} is apparent in Eq.~\ref{eq:leadlog} and has been regulated ``by hand'' at the scale $m_{D}$.  The ultraviolet regulator $k_{max}$ is the same as before.  The logarithmic dependence on energy (via $k_{max}$) is characteristic of collisional energy loss and is maintained even in the full calculation when one does not make the leading logarithm approximation.

The full calculation can be compared to the leading log approximations to the collisional energy loss that have been used previously in the literature.
\begin{equation}
\frac{dE_{A}}{dx}=-\frac{C_{F}\alpha_{s}}{4}m_{D}^{2}f(v)\log(B_{A})
\label{eq:litenloss}
\end{equation}
The index $A$ runs over the different calculations attempted.  They all agree with each other up to the kinematic ratio inside the logarithm $B_{A}=\frac{k_{max}}{k_{min}}$, given in \cite{wicks}.  We denote them as Bjorken (Bj) \cite{bjorken}, Thoma-Gyulassy (TG) \cite{tg} and Braaten-Thoma (BT) \cite{bt}.
\bea
%B_c= \left\{ 
%\begin{array}{ll}
B_{\rm Bj}&= & \hspace{0.15in}\left( 4ET \right)/\left(m_{D}^2\right)  \nonumber \\[1ex]
B_{\rm TG} &=& \hspace{0.15in}\left( \frac{4pT}{(E-p+4T)} \right) / \left(m_{D}\right) \nonumber\\[1ex]
B_{\rm BT} &=& \left\{ \begin{array}{ll}
% \hspace{0.15in}
\left(2^{\frac{n_f}{6+ n_f}}\; 
    0.85 \;E T \right)/ \left(\frac{m_{D}^2}{3}\right) & %\;\;\; \;,\;\;\; 
E \gg \frac{M^2}{T} \\[2ex] %\nonumber \\[1ex]
%B_{\rm BT} &=& \hspace{0.15in}
\left(4^{\frac{n_f}{6+ n_f}}\; 
    0.36 \; \frac{(E T)^2}{M^2} \right)/ \left(\frac{m_{D}^2}{3}\right) & %\;,\; 
E\ll \frac{M^2}{T} 
\end{array}\right.
\nonumber \\[1ex]
%\end{array}
%\right.
\label{eq:litB}
\eea
For purposes of calculation, the turnover in $B_{BT}$ is set at $E=2.36\frac{M^{2}}{T}$;  this value is determined by the condition that $B_{BT}$ should be a continuous function of $E$.

Figure \ref{fig:infonedeltE} shows the energy loss fraction $-\frac{\Delta E}{E}$ as a function of the initial momentum of the incident particle with the medium length traversed $L=5$ fm.  Curves are shown for the full calculation of $dE/dx$ as in Eq.~\ref{eq:preleadlog} and for all the previous leading log formulae for collisional energy loss detailed in Eqs.~\ref{eq:litenloss} and \ref{eq:litB}.  We include curves for the full calculation at two different values of $k_{max}$.  There is some ambiguity in setting the value of the momentum of the medium particle $q \sim T - 3T$.  Both curves are shown as a measure of the maximal theoretical uncertainty.  Note that all the calculations agree for asymptotically large energies (as they should) and that the full calculation (Eq.~\ref{eq:preleadlog}) does not possess the unphysical energy gain at low momentum that is shown by all the leading log approximations.

The agreement between the full calculation in this paper and the earlier leading log calculations is a non trivial result.  In particular, the Bjorken and Braaten-Thoma calculations are attempted in an entirely different formalism than our own.  Their formalism calculates collisional energy loss as an average over the quantum mechanical cross section for elastic collisions in the medium.  The formalism of the current paper and in Thoma-Gyulassy, however, uses linear response arguments in the abelianized classical Yang-Mills theory.

The agreement between our calculation and Thoma-Gyulassy is also a subtle one.  One can explicitly see by comparing Eqs.~\ref{eq:leadlog} and \ref{eq:litenloss} that our calculation is a factor of 2 larger than Thoma-Gyulassy, modulo different kinematic ratios in the logarithm.  This factor of two (also noticed by Peigne {\it et al.} \cite{peigne}) comes from the different polarization functions $\Pi_{\perp}(\omega,\vk)$ and $\Pi_{\|}(\omega,\vk)$ (defined in Appendix \ref{App1}) used in this paper as compared to the ones used in Thoma-Gyulassy.  Thoma-Gyulassy use only the positive frequency half ($\omega>0$) of the logarithmic branch cut present in the polarization function while the current paper takes into account the full branch cut extending across $-k<\omega<k$.  Thus, the Thoma-Gyulassy result is suppressed by a factor of 2.  

Their result is enhanced by the use of a large $k_{max}$ in the kinematic logarithm as compared to all other calculations.  They evaluate the maximum momentum transfer in an elastic collision in the plasma frame rather than in the center of mass frame, thus including a $\gamma$ factor as an enhancement in their result.  This enhancement can easily be seen by calculating the asymptotic energy dependence of $k_{max}$ in each model.  The Thoma-Gyulassy kinematic factor for asymptotic energies gives $k_{max}\sim E$ while all others give $k_{max}\sim\sqrt{ET}$.  We also note the calculation by Djordjevic \cite{djordj} uses the same kinematics and branch cut characteristics as Thoma-Gyulassy and thus the momentum dependence of energy loss in the Djordjevic model is very similar to the Thoma-Gyulassy model, despite the fact that in spirit and calculational formalism Djordjevic is closer to Bjorken or Braaten-Thoma.

\subsection{Infinite Time Two Particle Current}

We now employ the two particle infinite current as in Eq.~\ref{eq:infv20} in order to clarify the effects of binding without the complication of a finite time analysis.  The end result of the finite time analysis will remove all reference to this binding effect.  We do the calculations in this section on our way to make contact with the calculation of Peigne {\it et al.} \cite{peigne}, showing how including this unrealistically enhanced contribution leads to a suppression in the energy loss. 

The current in Eq.~\ref{eq:infv20} includes components of the energy loss that can be thought of as a self interaction (equivalent to the one particle case) while the rest is a shared interaction between the stationary second particle and the moving particle in question.  We will call terms of the first type {\it self} terms while the second kind will be {\it shared} terms.

In order to facilitate easy physical interpretation of the different sources of the energy loss, it is convenient to decompose the retarded propagator in terms of a purely advanced propagator $D_{A}$, and a difference of the retarded and advanced propagator $D_{-}$, i.e.  $D_{R}=D_{R}-D_{A}+D_{A}=D_{-}+D_{A}$.  The procedure is detailed in Appendix \ref{App3}.  This helps to isolate the singularities in the complex $\omega$ plane that are present in the propagator and thus manifestly breaks up the expression in Eq. \ref{eq:dEdx} into contributions from different sources.  This is particularly useful for the finite time current to be studied later and will be used in this section for purely illustrative purposes.

We will calculate the vacuum and medium field energy losses separately, paying close attention to which parts of the energy loss are coming from true feedback in the plasma and which are due to the interactions with the stationary particle.

\subsubsection{Vacuum Field Contributions}

We start by looking at terms in the energy loss that come from the vacuum field specified in Eq.~\ref{eq:fields}.  The expression for the retarded propagator $D_{R}(\omega,\vk)$ can be taken from Appendix \ref{App1} and its decomposition into $D_{R}=D_{-}+D_{A}$ is detailed in Appendix \ref{App3}.  The energy loss from fields due to the $D_{-}(\omega,\vk)$ propagator are as follows.
\begin{eqnarray}
\frac{dE _{vac}^{-}}{dx}=\int \invvol e^{-i(\omega-kv\mu)\frac{L}{v}} \times  \qquad \qquad \nonumber \\
D_{-}(\omega,\vk)(\omega \vj^{a}_{\infty,(2)}(\omega,\vk) -\vk j^{0 a}_{\infty,(2)}(\omega,\vk))
\end{eqnarray}
As explained earlier, this contribution can be broken up into a self interaction and a shared interaction.  Using Eq. \ref{eq:infv20} we write $\omega \vj^{a}_{\infty,(2)} -\vk j^{0 a}_{\infty,(2)}=2\pi q^{a}((\omega\vv{}-\vk)\delta(\omega-\vk\cdot\vv{})+\vk\delta(\omega))$.  The part proportional to $\delta(\omega-\vk\cdot\vv{})$ contributes to the self interaction while everything else is a shared interaction.  We separate them,
\begin{eqnarray}
\frac{dE_{vac}^{-,self}}{dx}&=&\frac{C_{R}\alpha_{s}}{v}(1-v^{2})\int  { } k\mu \{ \delta\left(\mu-\frac{1}{v}\right)\nonumber \\
& & { } -\delta\left(\mu+\frac{1}{v}\right)\}dkd\mu \nonumber \\
&=&0\nonumber \\
\frac{dE_{vac}^{-,shar}}{dx}&=&-C_{R}\alpha_{s}\int k^{2}\mu e^{ik\mu L} ( \delta(-k)-\delta(k))dkd\mu\nonumber \\
&=&0
\end{eqnarray}
As shown in Appendix \ref{App3}, the vacuum $D_{-}$ contribution is just twice the radiative pole contribution.  For an infinite time current, the only kind of radiation possible is Cerenkov radiation.  The characteristic Cerenkov cone is present in the self interaction expression as $\mu=\cos(\theta)=\frac{1}{v}$.  This contribution goes to zero as the velocity of the particle is less than that of light.

The only contribution left over is that due to the advanced propagator $D_{A}(\omega,\vk)$.  This can be written as,
\begin{eqnarray}
\frac{dE_{vac}^{Adv}}{dx}=\frac{q^{a}}{v}\int \invvol e^{-i(\omega-kv\mu)\frac{L}{v}} \times  \qquad \qquad \nonumber \\
D_{A}(\omega,\vk)(\omega \vv{}\cdot\vj^{a}_{\infty,(2)}(\omega,\vk) -\vv{}\cdot\vk j^{0 a}_{\infty,(2)}(\omega,\vk))
\label{advanexpr}
\end{eqnarray}
The self interaction part integrates to zero but there is a non zero shared interaction.
\begin{eqnarray}
\frac{dE_{vac}^{Adv,shar}}{dx}= C_{F}\alpha_{s}\int \frac{d^{3}\vk}{(2\pi)^{3}}\hat{\vv{}}\cdot\vk e^{ik\mu L}D_{A}(0,\vk)
\end{eqnarray}

This vacuum shared interaction is a measure of the binding energy of the pair of particles and is the only source of vacuum energy loss for the moving particle.  It can be thought of as the energy lost by the moving particle as it climbs out of the binding potential well of the stationary particle of opposite charge.  This integral is nominally infinite as both particles appear at the origin at $t=0$.  The divergence is simply due to the infinite shared energy of two point charges that goes like $k_{max}\sim \frac{1}{r_{max}}$.  It is typical of the binding (shared) terms seen in the analysis and should be regulated using the length scale $|\vbb|$ (and hence momentum scale $Q$) of the initial hard interaction.  In order to be consistent with our simplified current which has no information about an initial hard interaction and with the medium contributions that have their own prescribed interaction scale, we will use the previously defined medium $k_{max}$ as our regulator.

\subsubsection{Medium Field Contributions}

The medium field can be conveniently broken up into a transverse and a longitudinal contribution and those individually can be further subdivided into a $D_{-}(\omega,\vk)$  contribution and a $D_{A}(\omega,\vk)$.  As shown in Appendix $\ref{App3}$, this decomposition leads to energy loss contributions from three different physical processes: the particle-hole cut, the plasmon radiation loss, and the advanced coulombic field.

We begin by considering the longitudinal electric field in the medium (as given by Eq.~\ref{eq:fields}).  As in the vacuum case, the only kind of plasmon radiation possible is of Cerenkov type, which is not present as $\omega_{pl}^{\|}(k)\ge k$ (as shown in Appendix \ref{App3}).  We therefore move on to contributions from the particle-hole branch cut.
\begin{eqnarray}
\frac{dE_{L}^{cut,self}}{dx}=C_{F}\alpha_{s}\int \frac{d^{3}\vk}{(2\pi)^{3}}  \hat{\vv{}}\cdot\hat{\vk} 
A^{\|}(\vk\cdot\vv{},\vk)
\end{eqnarray}
The function $A^{\|}(\omega,\vk)$ is defined in Appendix $\ref{App3}$.  It is the part of $D_{-}^{\|}(\omega,\vk)$ that is linked back to the logarithmic particle-hole branch cut.  Thus, it only has support in frequency for $-k<\omega<k$.  The shared contribution to the particle hole cut integrates to zero since $A^{\|}(0,\vk)=0$.

There also remains an advanced longitudinal field that causes energy loss.  This contribution can be written as,
\begin{eqnarray}
\frac{dE_{\|}^{Adv,self}}{dx}=C_{F}\alpha_{s}\int \frac{d^{3}\vk}{(2\pi)^{3}}  \hat{\vv{}}\cdot\hat{\vk} 
D_{A}^{\|}(\vk\cdot\vv{},\vk) \nonumber \\
\frac{dE_{\|}^{Adv,shar}}{dx}=  -C_{F}\alpha_{s}\int \frac{d^{3}\vk}{(2\pi)^{3}}\hat{\vv{}}\cdot\vk e^{i\vk\cdot\hat{\vv{}}L}D_{A}^{\|}(0,\vk).
\end{eqnarray}
$D_{A}^{\|}(\omega,\vk)$ is the longitudinal in medium propagator with advanced boundary conditions.

Due to the form of the current $\vj^{a}_{\infty,(2)}(\omega,\vk)$, the transverse field only provides self interactions and no shared interactions.  As in the longitudinal medium and the full vacuum cases, the radiative part of the field has to be purely Cerenkov in nature and thus does not contribute.  We begin with the transverse particle-hole branch cut,
\begin{equation}
\frac{dE_{\perp}^{cut,self}}{dx}=C_{F}\alpha_{s}v\int\!\! \frac{d^{3}\vk}{(2\pi)^{3}} (1\!\!- \!\!(\hat{\vv{}}\!\cdot\!\hat{\vk})^{2}) A^{\perp}(\vk\!\cdot\!\vv{},\vk)
\end{equation}
$A^{\perp}(\omega,\vk)$ is defined in Appendix \ref{App3} and is the part of the $D_{-}^{\perp}(\omega,\vk)$ propagator that can be associated with the logarithmic branch cut singularity structure of the particle-hole continuum.  Similar to the longitudinal case, this function only has support in frequency for $-k<\omega<k$.  The only remaining piece is the advanced transverse field.
\begin{equation}
\frac{dE_{\perp}^{Adv,self}}{dx}=C_{F}\alpha_{s}v\int\!\! \frac{d^{3}\vk}{(2\pi)^{3}} (1\!\!- \!\!(\hat{\vv{}}\!\cdot\!\hat{\vk})^{2}) D_{A}^{\perp}(\vk\!\cdot\!\vv{},\vk)
\end{equation}
$D_{A}^{\perp}(\omega,\vk)$ is the transverse in medium propagator with advanced boundary conditions.

\subsubsection{Total $dE/dx$}

\begin{figure}
\centering
 \epsfig{file=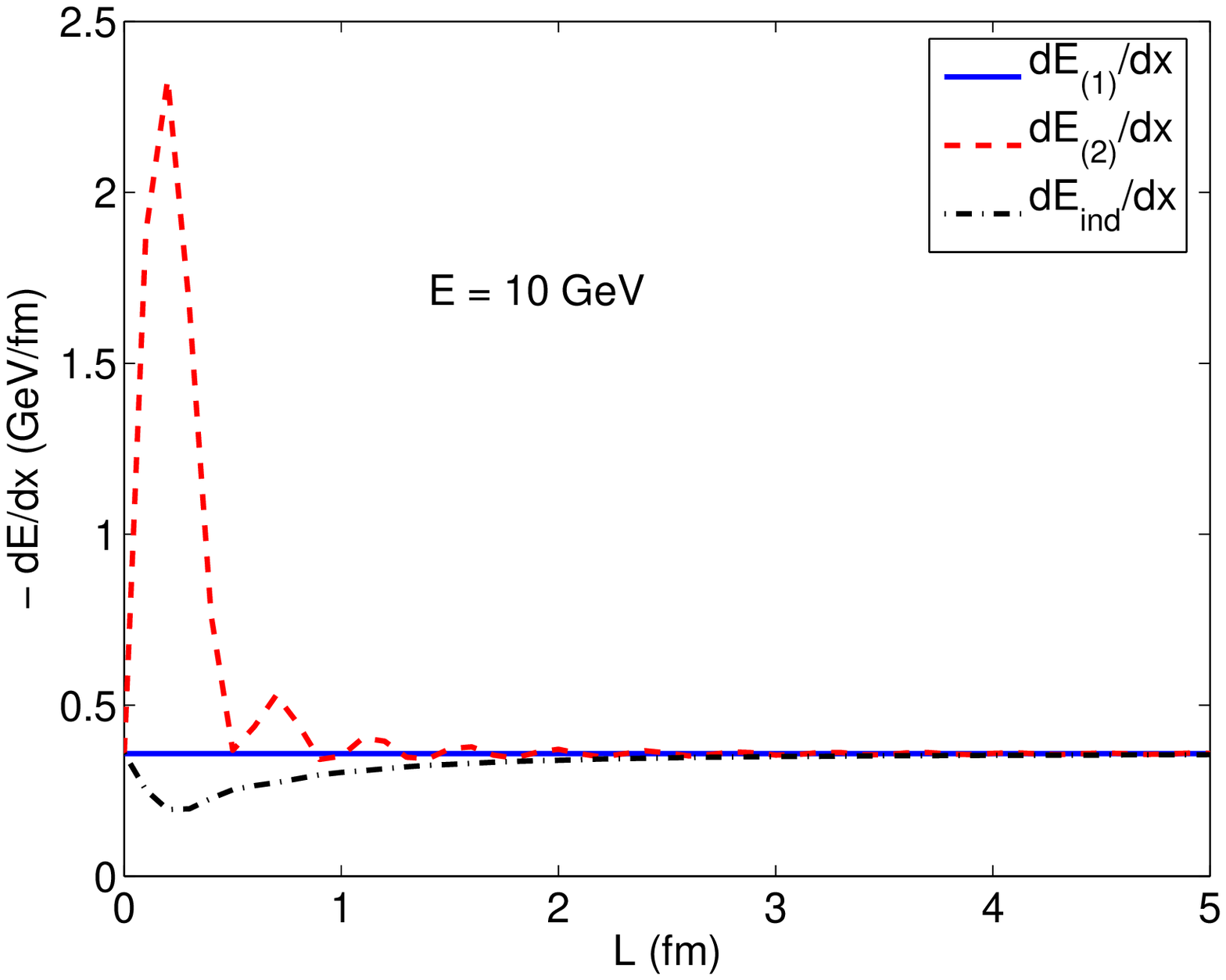,width=3.5in,angle=0}
  \caption{(Color Online) The figure shows the differential energy loss  $-dE/dx$ as a function of the path length traversed in the medium $L$.  Curves are shown for an incident charm quark with $M=1.2$ GeV and initial energy $E=10$ GeV.}
 \label{fig:inf2partdEdx}
\end{figure}

\begin{figure}
\centering
 \epsfig{file=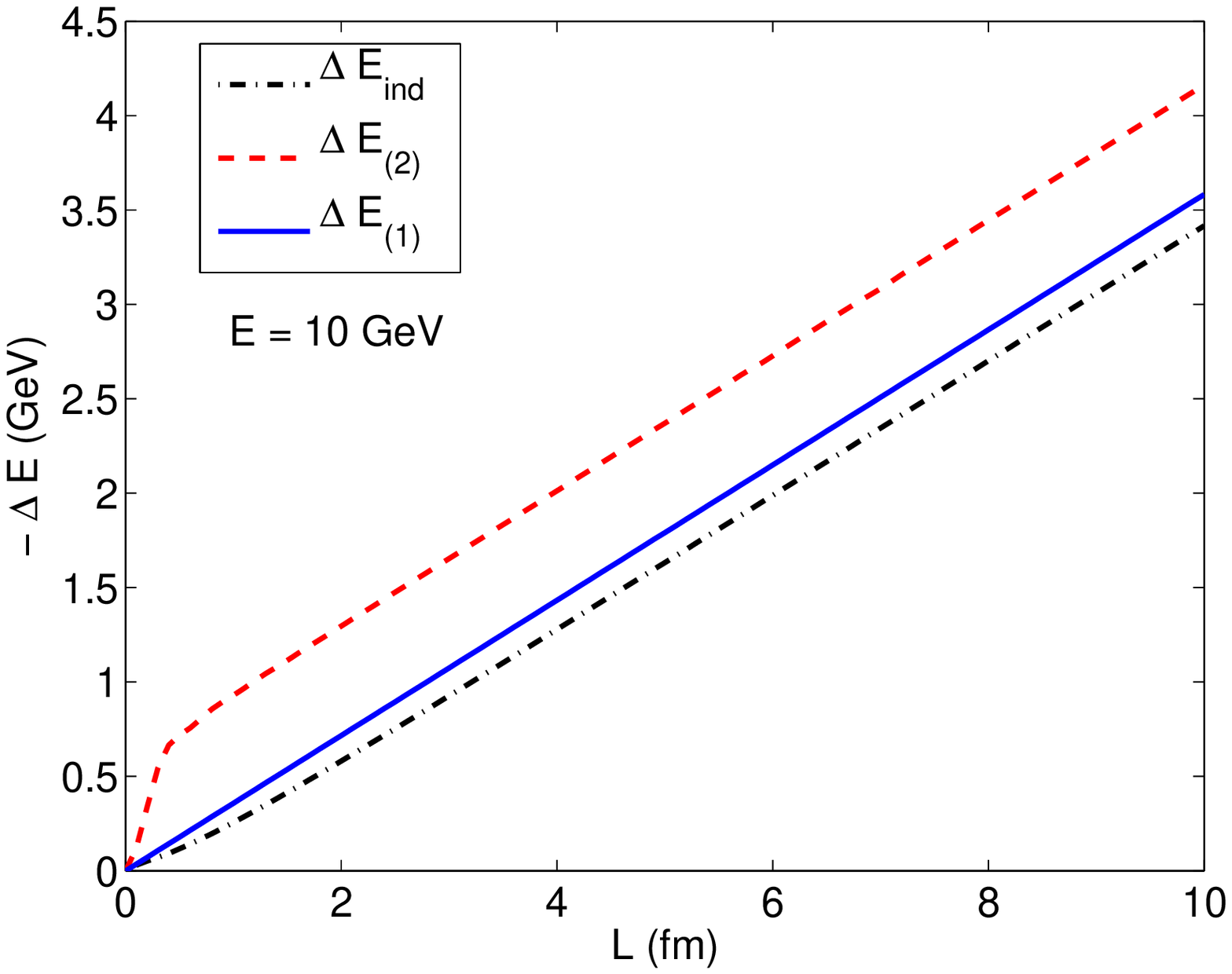,width=3.5in,angle=0}
  \caption{(Color Online) The figure shows the energy loss  $-\Delta E$ as a function of the path length traversed in the medium $L$.  Curves are shown for an incident charm quark with $M=1.2$ GeV and initial energy $E=10$ GeV.}
 \label{fig:inf2partdelE}
\end{figure}

\begin{figure}
\centering
 \epsfig{file=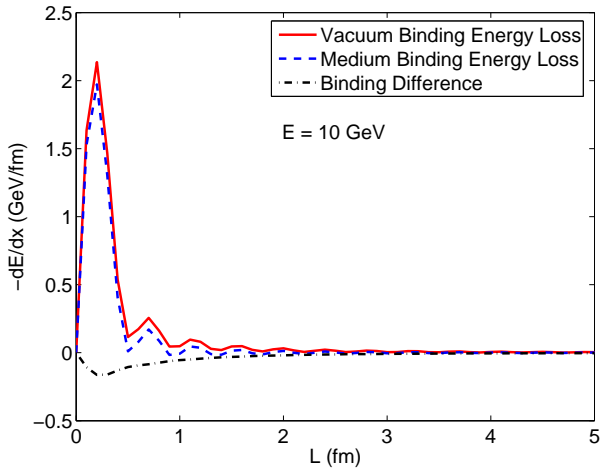,width=3.5in,angle=0}
  \caption{(Color Online) The figure shows the differential energy loss  $-dE/dx$ due to the binding between the two particles that constitute the incident current as a function of the path length traversed in the medium $L$.  Curves are shown for an incident charm quark with $M=1.2$ GeV and initial energy $E=10$ GeV.}
 \label{fig:inf2partbind}
\end{figure}

We now add up all the energy loss due to the medium fields and subtract all the vacuum energy loss expressions in order to determine the total work done by the induced field.
\begin{equation}
\frac{dE_{ind}}{dx}=\sum_{A}\frac{dE_{med}^{A}}{dx}-\sum_{B}\frac{dE_{vac}^{B}}{dx}
\label{eq:totdEdx2part}
\end{equation}
The indices $A,B$ run over the different types of energy loss (e.g. from the branch cut, from advanced fields), both self and shared.  Figs.~\ref{fig:inf2partdEdx} and \ref{fig:inf2partdelE} show this energy loss.  The black dot-dashed curve in the figures shows the differential energy loss ($dE_{ind}/dx$) as calculated in Eq.~\ref{eq:totdEdx2part}.  The blue solid curve is the full calculation for the infinite time one particle energy loss as shown earlier in the paper.  Note that the induced field energy loss (in Fig.~\ref{fig:inf2partdEdx}) is actually smaller for some time than the one particle energy loss. 

The initial dip in $dE_{ind}/dx$ (seen in both Figs.~\ref{fig:inf2partdEdx} and \ref{fig:inf2partdelE}) is caused by the subtraction of the vacuum contribution from the medium parts of the energy loss as stated in the induced field prescription.  Why this should lead to a dip in energy loss is seen in Fig.~\ref{fig:inf2partbind}.  Here we plot only the shared (binding) terms that contribute to the problem.  Due to screening effects, the in medium energy loss is less than the vacuum energy loss.  In our calculation for $dE_{ind}/dx$ in Eq.~\ref{eq:totdEdx2part} we have taken a difference between these binding energies causing a small energy gain; a ``Ter-Mikaelian'' adjustment to the binding energy.  This is exactly the effect that is included in Peigne {\it et al.} \cite{peigne}; an energy gain from the difference of binding energies between vacuum and medium interactions.  The magnitude of this component is also enhanced because of the unrealistic charge correlation between the two particles in the current and by the fact that the second particle is created at rest at the origin.

In order to make sure that this depression in the energy loss is coming purely from the difference of shared interaction we can systematically remove the effect.
\begin{eqnarray}
\frac{dE_{(2)}}{dx}=\sum_{A}\frac{dE_{med}^{A}}{dx} \nonumber \\
\frac{dE_{(1)}}{dx}=\sum_{A\in\{self\}}\frac{dE_{med}^{A}}{dx}
\end{eqnarray}
If we leave out only the vacuum interactions but keep all the two particle interactions in the medium we get $dE_{(2)}/dx$ shown as the red dashed curve in Figs. \ref{fig:inf2partdEdx} and \ref{fig:inf2partdelE}.  This result makes sense as it is just the one particle infinite result with the additional energy loss due to binding between the two particles in the medium.  We attain $dE_{(1)}/dx$ by removing all mention of the shared interaction entirely.  This result goes back to the infinite time one particle case, as it should.  This exercise is a good proof in principle that we can perform the necessary subtractions in the finite time case in order to isolate the effect we want to study.

\section{The Finite Time Problem}\label{Sec3}
The subtleties involved in choosing a conserved current for the finite time case are manifest in this section, which means that we must start with the two particle finite time current $j^{\mu a}_{(2)}(\omega,\vk)$.  We will analyze the full energy loss, specifying exactly how to remove the effects of the second particle which cause the anomalously small results of Peigne {\it et. al.} \cite{peigne} as well as the radiative parts that are included in our energy loss calculation (as shown in Appendix \ref{App2}).  Note that we will not combine the two separate terms inside $j^{0a}_{(2)}(\omega,\vk)$ into one term but rather will treat each term distinctly as a self and shared interaction.

\subsection{Induced Field Energy Loss and Removing Binding Effects}\label{inducedsec}

\subsubsection{Vacuum Field Contributions}

We begin the analysis by considering the contribution from the advanced propagator $D_{A}(\omega,\vk)$.  The analysis is greatly simplified by the fact that all the singularities in the propagator are in the upper half of the complex $\omega$ plane while the current singularities all reside in the lower half.  We can thus perform a contour integral in frequency space and close the contour in the lower half plane, picking up the residues at the current singularities.  This calculation leads to the same advanced field energy loss as in the infinite time case (Eq.~\ref{advanexpr}), as the current pole has the same frequency dependence as the infinite time current.  This will be true of all the advanced contributions in the paper.
\begin{eqnarray}
\frac{dE_{vac}^{Adv,shar}}{dx}=  C_{F}\alpha_{s}\int \frac{d^{3}\vk}{(2\pi)^{3}}\hat{\vv{}}\cdot\vk e^{ik\mu L}D_{A}(0,\vk)
\end{eqnarray}

We can now move on to the $D_{-}(\omega,\vk)$ contribution.  The self and shared contributions are,
\begin{eqnarray}
\frac{dE_{vac}^{-,self}}{dx}&=&-i\frac{C_{F}\alpha_{s}}{v (2\pi)^{2}}\int d^{3}\vk(\Gamma^{+}_{self}+\Gamma^{-}_{self})\nonumber \\
\Gamma^{+}_{self}&=&\frac{e^{-i(k-kv \mu)\frac{L}{v}}}{k}\frac{kv^{2}-\vv{}\cdot\vk}{k-\vk\cdot\vv{}} \nonumber \\
\Gamma^{-}_{self}&=&-\frac{e^{i(k+kv\mu)\frac{L}{v}}}{k}\frac{kv^{2}+\vv{}\cdot\vk}{k+\vk\cdot\vv{}}
\end{eqnarray}
\begin{eqnarray}
\frac{dE_{vac}^{-,shar}}{dx}&=&-i\frac{C_{F}\alpha_{s}}{v (2\pi)^{2}}\int d^{3}\vk(\Gamma^{+}_{shar}+\Gamma^{-}_{shar})\nonumber \\
\Gamma^{+}_{shar}&=&\frac{e^{-i(k-kv\mu)\frac{L}{v}}}{k}\frac{\vv{}\cdot\vk}{k} \nonumber \\
\Gamma^{-}_{shar}&=&\frac{e^{i(k+kv\mu)\frac{L}{v}}}{k}\frac{\vv{}\cdot\vk}{k}.
\end{eqnarray}
 Unlike the infinite case, there will actually be a non zero contribution from the $D_{-}(\omega,\vk)$ propagator.  This is easily interpreted as the radiation emitted when a charge is created out of the vacuum.
 
\subsubsection{Medium Field Contributions}
The contribution from the longitudinal particle hole cut is,
\begin{eqnarray}
\frac{dE_{\|}^{cut,self}}{dx}=iC_{F}\alpha_{s}\!\!\int \!\!\invvol e^{-i(\omega-kv\mu)\frac{L}{v}} \frac{A^{\|}(\omega,\vk)\hat{\vv{}}\cdot\hat{\vk}}{\omega-\vk\cdot\vv{}+i\eta} \nonumber \\
\frac{dE_{\|}^{cut,shar}}{dx}=-iC_{F}\alpha_{s}\!\!\int\!\! \invvol e^{-i(\omega-kv\mu)\frac{L}{v}} \frac{A^{\|}(\omega,\vk)\hat{\vv{}}\cdot\hat{\vk}}{\omega+i\eta}
\end{eqnarray}
The integrals in the previous expressions can be peformed by doing the angular integrals first in terms of the Exponential Integral function.  The $D_{A}^{\|}(\omega,\vk)$ contribution is easy to calculate via contour integration in the $\omega$ plane.
\begin{eqnarray}
\frac{dE_{\|}^{Adv,self}}{dx}=C_{F}\alpha_{s}\int \frac{d^{3}\vk}{(2\pi)^{3}}  \hat{\vv{}}\cdot\hat{\vk} 
D_{A}^{\|}(\vk\cdot\vv{},\vk) \nonumber \\
\frac{dE_{\|}^{Adv,shar}}{dx}=  -C_{F}\alpha_{s}\int \frac{d^{3}\vk}{(2\pi)^{3}}\hat{\vv{}}\cdot\vk e^{ik\mu L}D_{A}^{\|}(0,\vk)
\end{eqnarray}
There is now a non zero contribution in the finite time case from the plasmon term inside the $D_{-}^{\|}(\omega,\vk)$  propagator.  It can be written down as,
\begin{eqnarray}
\frac{dE_{\|}^{pl,self}}{dx}&=&-i\frac{2C_{F}\alpha_{s}}{v (2\pi)^{2}}\int d^{3}\vk(\Gamma^{\|,+}_{self}+\Gamma^{\|,-}_{self})\nonumber \\
\Gamma^{\|,+}_{self}&=&\frac{e^{-i(\omega_{pl}^{\|}-kv\mu)\frac{L}{v}}}{f(\omega_{pl}^{\|},\vk)}\frac{\vv{}\cdot\vk}{\omega_{pl}^{\|}-\vk\cdot\vv{}} \nonumber \\
\Gamma^{\|,-}_{self}&=&-\frac{e^{i(\omega_{pl}^{\|}+kv\mu)\frac{L}{v}}}{f(-\omega_{pl}^{\|},\vk)}\frac{\vv{}\cdot\vk}{\omega_{pl}^{\|}+\vk\cdot\vv{}}
\end{eqnarray}
\begin{eqnarray}
\frac{dE_{\|}^{pl,shar}}{dx}&=&i\frac{2C_{F}\alpha_{s}}{v (2\pi)^{2}}\int d^{3}\vk(\Gamma^{\|,+}_{shar}+\Gamma^{\|,-}_{shar})\nonumber \\
\Gamma^{\|,+}_{shar}&=&\frac{e^{-i(\omega_{pl}^{\|}-kv\mu)\frac{L}{v}}}{f(\omega_{pl}^{\|},\vk)}\frac{\vv{}\cdot\vk}{\omega_{pl}^{\|}} \nonumber \\
\Gamma^{\|,-}_{shar}&=&-\frac{e^{i(\omega_{pl}^{\|}+kv\mu)\frac{L}{v}}}{f(-\omega_{pl}^{\|},\vk)}\frac{\vv{}\cdot\vk}{\omega_{pl}^{\|}}
\end{eqnarray}
These plasmon contribution are those field components that can support the longitudinal plasmon dispersion relation relation $\omega_{pl}^{\|}(k)$, plotted in Appendix \ref{App2}.  The function $f(\omega,\vk)$ is the longitudinal plasmon wave function renormalization factor and is defined in Appendix \ref{App2} in terms of the derivative of the polarization function $\Pi_{\|}(\omega,\vk)$.

There are no shared contributions from the transverse fields, only self interactions.  Contributions from the transverse particle-hole cut can be written as,
\begin{eqnarray}
\frac{dE_{\perp}^{cut,self}}{dx}=iC_{F}\alpha_{s}v \int \invvol e^{-i(\omega-kv\mu)\frac{L}{v}} \nonumber \\
\times \frac{A^{\perp}(\omega,\vk)(1-(\hat{\vv{}}\cdot\hat{\vk})^{2})}{\omega-\vk\cdot\vv{}+i\eta}.
\end{eqnarray}
We can also get the advanced contribution very easily from contour integration.
\begin{eqnarray}
\frac{dE_{\perp}^{Adv,self}}{dx}=C_{F}\alpha_{s}v\int\!\! \frac{d^{3}\vk}{(2\pi)^{3}} (1\!\!- \!\!(\hat{\vv{}}\!\cdot\!\hat{\vk})^{2}) D_{A}^{\perp}(\vk\!\cdot\!\vv{},\vk)
\end{eqnarray}
The plasmon terms coming from the transverse parts of the field can be written as,
\begin{eqnarray}
\frac{dE_{\perp}^{pl,self}}{dx}&=&-i\frac{2C_{F}\alpha_{s}}{v (2\pi)^{2}}\int d^{3}\vk(\Gamma^{\perp,+}_{self}+\Gamma^{\perp,-}_{self})\nonumber \\
\Gamma^{\perp,+}_{self}&=&\frac{\omega_{pl}^{\perp}e^{-i(\omega_{pl}^{\perp}-kv\mu)\frac{L}{v}}}{g(\omega_{pl}^{\perp},\vk)}\frac{1-(\vv{}\cdot\vk)^{2}}{\omega_{pl}^{\perp}-\vk\cdot\vv{}} \nonumber \\
\Gamma^{\perp,-}_{self}&=&\frac{\omega_{pl}^{\perp}e^{i(\omega_{pl}^{L}+kv\mu)\frac{L}{v}}}{g(-\omega_{pl}^{\perp},\vk)}\frac{1-(\vv{}\cdot\vk)^{2}}{\omega_{pl}^{\perp}+\vk\cdot\vv{}}
\end{eqnarray}
The function $g(\omega,\vk)$ is the transverse plasmon wave function normalization factor and is defined in Appendix \ref{App2} in terms of the derivative of the polarization function $\Pi_{\perp}(\omega,\vk)$.

\subsubsection{Total $\frac{dE}{dx}$}

\begin{figure}
\centering
 \epsfig{file=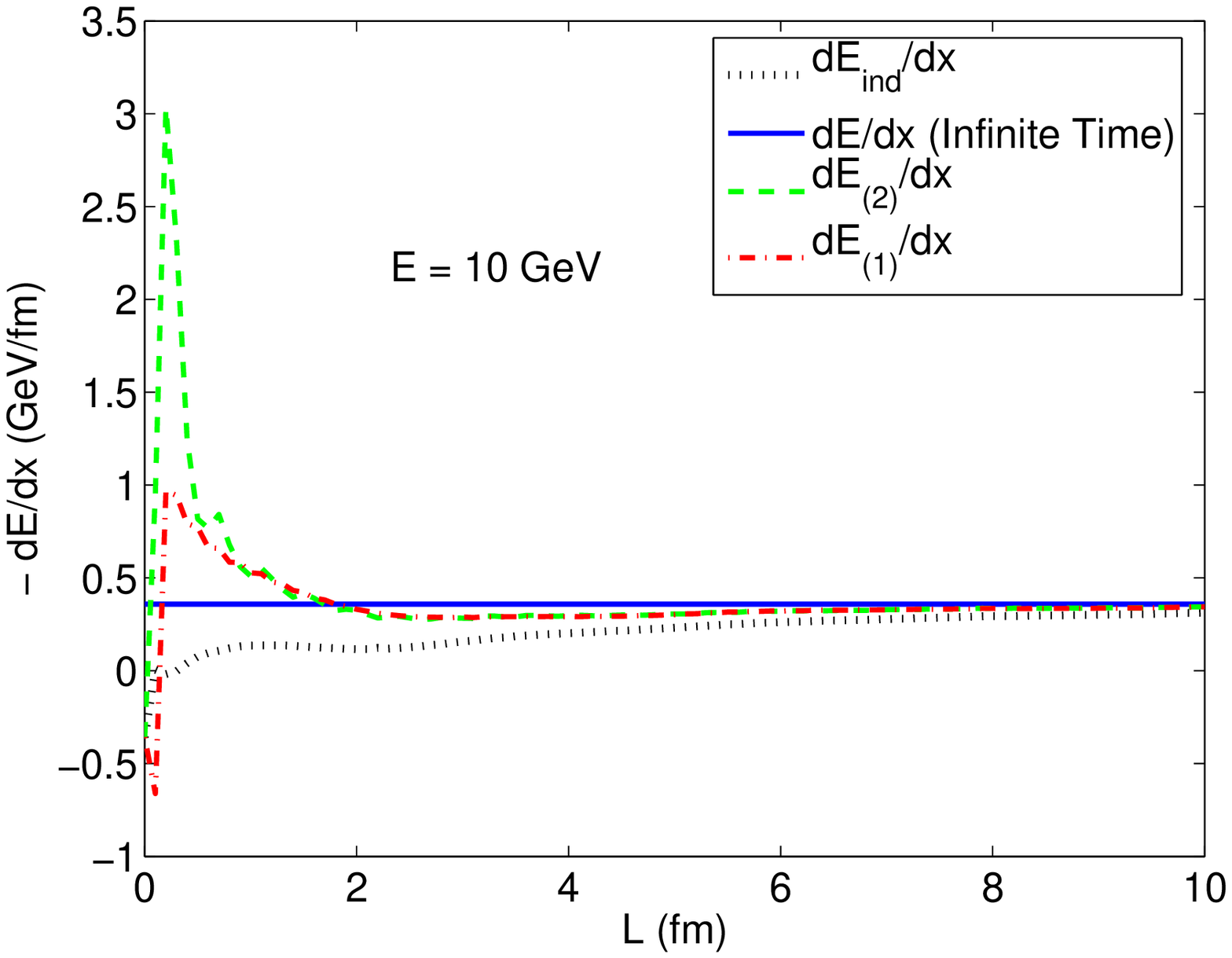,width=3.5in,angle=0}
  \caption{(Color Online) The figure shows the differential energy loss  $-dE/dx$ as a function of the path length traversed in the medium $L$.  Curves are shown for an incident charm quark with $M=1.2$ GeV and initial energy $E=10$ GeV and include radiative modes.}
 \label{fig:fin2partdEdx}
\end{figure}

\begin{figure}
\centering
 \epsfig{file=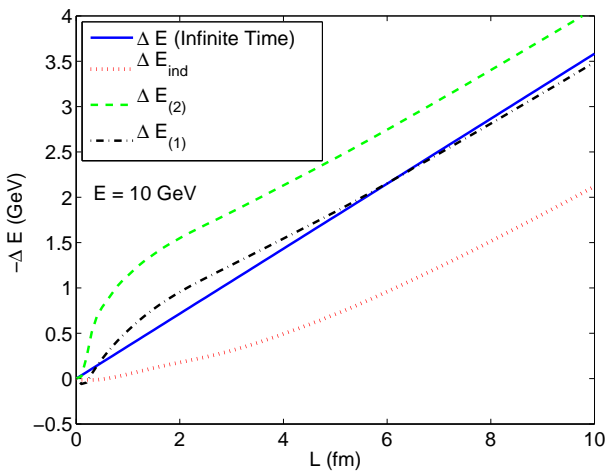,width=3.5in,angle=0}
  \caption{(Color Online) The figure shows the energy loss  $-\Delta E$ as a function of the path length traversed in the medium $L$.  Curves are shown for an incident charm quark with $M=1.2$ GeV and initial energy $E=10$ GeV and include radiative modes.}
 \label{fig:fin2partdelE}
\end{figure}

The first object we calculate is the energy loss due to the total induced field, making no distinction as to whether the field is coming from a self or shared interaction.
\begin{equation}
\frac{dE_{ind}}{dx}=\sum_{A}\frac{dE_{med}^{A}}{dx}-\sum_{B}\frac{dE_{vac}^{B}}{dx}
\end{equation}
The indices $A$ and $B$ sum over all the different sources of energy loss, both self and shared.  The results are shown in Figs.~\ref{fig:fin2partdEdx} and \ref{fig:fin2partdelE}.  The solid blue curve in the figures is the answer to the infinite time single particle problem, presented for comparison.  The total induced field energy loss (counting both self and shared interactions) is shown as the dotted line.  Note that the induced field energy loss is far less than the infinite case and does not get back up to the infinite answer (in Fig.~\ref{fig:fin2partdelE}) even after traveling up to 10 fm.  This calculation reproduces Peigne {\it et al.} \cite{peigne} and is different only in as much as the parameter sets used are different.

As explained in the previous section, a naive calculation of energy loss from the total induced electric field includes shared interactions as well as the self ones.  The shared contribution is unrealistic and needs to be systematically removed from the calculation.  We first remove the vacuum shared field,
\begin{equation}
\frac{dE_{(2)}}{dx}=\sum_{A}\frac{dE_{med}^{A}}{dx}-\sum_{B \in \{ self \}}\frac{dE_{vac}^{B}}{dx}
\end{equation}
We refer to the energy loss that neglects the binding difference as $dE_{(2)}/dx$ as it explicitly shows the effect of the stationary particle's in medium potential well on the moving particle.  The dashed curves in the Figs.~\ref{fig:fin2partdEdx} and \ref{fig:fin2partdelE} show this situation, which is just the one particle finite contribution plus the energy loss due to in medium binding.

In order to compare the finite time calculation with the equivalent infinite time answer, we need to remove all the shared parts (medium as well as vacuum).
\begin{equation}
\frac{dE_{(1)}}{dx}=\sum_{A \in \{ self \}}\frac{dE_{med}^{A}}{dx}-\sum_{B \in \{ self \}}\frac{dE_{vac}^{B}}{dx}
\end{equation}
The result of the purely self interactions is shown as the dot-dash curves in Figs.~\ref{fig:fin2partdEdx} and \ref{fig:fin2partdelE}.  The finite time calculation starts lower than the infinite time case as the induced field takes some time to build up and be created in the medium.  We can now see in Fig.~\ref{fig:fin2partdelE} that the dot dash-curve (which still includes radiative energy loss effects both gluonic and plasmonic) is getting closer to the infinite time case.  We have systematically removed the unphysical binding terms and shown that they were the cause of the large suppression of energy loss in Peigne {\it et al.} \cite{peigne}. 

\subsection{Subtracting the Radiative Ter-Mikaelian Effect}

\begin{figure}
\centering
 \epsfig{file=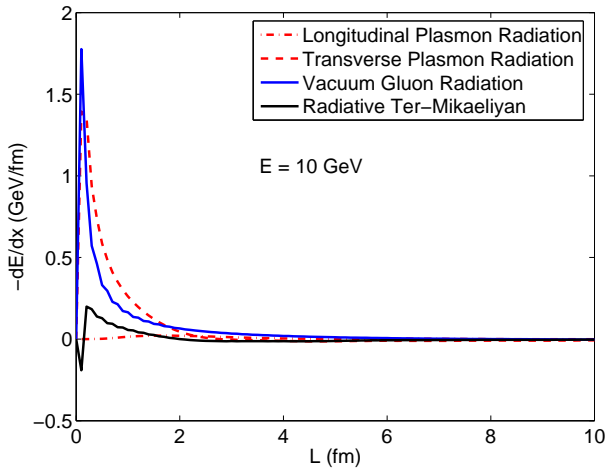,width=3.5in,angle=0}
  \caption{(Color Online) The figure shows the differential energy loss  $-dE/dx$ as a function of the distance traveled $L$ due to radiative effects at zeroth order in opacity.  Curves are shown for an incident charm quark with $M=1.2$ GeV and initial energy $E=10$ GeV.}
 \label{fig:termikdEdx}
\end{figure}

\begin{figure}
\centering
 \epsfig{file=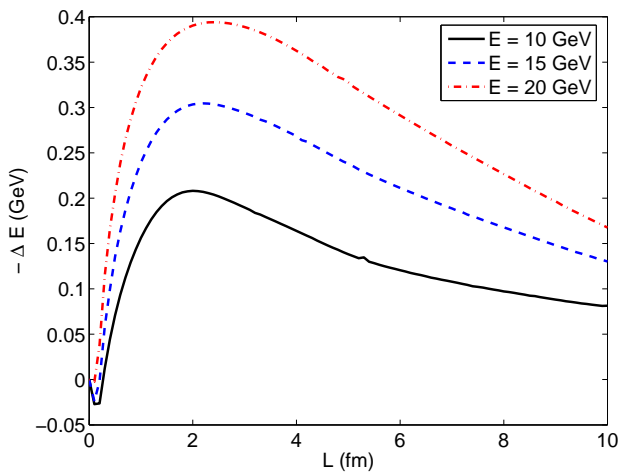,width=3.5in,angle=0}
  \caption{(Color Online) The figure shows the energy loss  $-\Delta E$ as a function of the distance traveled $L$ due to radiative effects at zeroth order in opacity.  Curves are shown for an incident charm quark with $M=1.2$ GeV and different initial energies $E=10,15,20$ GeV.}
 \label{fig:termikdelE}
\end{figure}

 We note that the one particle result quoted in the previous section includes more effects than just the pure collisional energy loss suffered by the particle.  When a current is created out of the vacuum, it carries with itself certain off shell Fourier modes that can be radiated off to infinity in the form of gluons in the vacuum and plasmons in the medium.  As shown in Appendix \ref{App2}, the Joule heating expression we use includes contributions that become the total zeroth order in opacity radiative energy loss when we integrate $dE/dx$ for an infinite path length.  This radiative energy loss needs to be removed from the calculation if we are to correctly compare the finite time result with previous estimates for the collisional energy loss.
 
The radiative effects manifest themselves in terms that can be schematically written down as $\frac{dE^{A}_{rad}}{dx}=D_{rad}j$, where the index $A$ can represent a vacuum, medium transverse, or medium longitudinal contribution and the function $D_{rad}(\omega,\vk)$ is the radiative Green function as specified in Appendix \ref{App2}.  The radiative energy loss curves for a particle with initial energy $E=10$ GeV in both vacuum (solid blue curve) and medium (dashed and dash-dotted curves) are shown in Fig. \ref{fig:termikdEdx}.  There is only transverse gluon radiation in the vacuum but the medium supports both transverse and longitudinal polarizations for the plasmons.  The longitudinal energy loss, however, is significantly smaller than the energy loss caused by transverse plasmons.  Note that the vacuum peak is larger than the transverse peak.
  
The Ter-Mikaelian effect (first studied by Djordjevic {\it et al.} \cite{termik}) is the difference between the zeroth order radiative energy loss in the medium and the zeroth order vacuum radiative energy loss (shown in Fig.~\ref{fig:termikdEdx}).  We implicitly include this difference in the calculations for $dE_{(1)}/dx$ at the end of the last section when we calculate the induced field ($\vE^{a}-\vE^{a}_{vac}$).  It is shown in Appendices \ref{App2} and \ref{App3} that the radiative part can be isolated by carefully dealing with the singularities in the complex $\omega$ plane.  The result for the radiative part is,
\begin{equation}
\frac{dE_{Rad}}{dx}=\frac{1}{2}\left( \sum_{A\in \{pl\}}\frac{dE^{A}_{med}}{dx} - \sum_{B\in \{-\}}\frac{dE^{B}_{vac}}{dx} \right)
\end{equation}
The factor of a half comes from the $D_{-}(\omega,\vk)$ propagator.  This difference propagator has a part that has the correct radiative dispersion relation but has a residue of twice the radiative residue in the retarded propagator.

The difference between medium and vacuum radiative energy loss at $E=10$ GeV is shown as the solid black curve in Fig. \ref{fig:termikdEdx}.  Note that this curve includes the effects of the longitudinal radiation (something which is neglected in Djordjevic {\it et al.} \cite{termik}) and also has a complete calculation of the wave function renormalization factor (Appendix \ref{App2}).  Fig. \ref{fig:termikdelE} shows the total Ter-Mikaelian energy radiated by the particle as a function of the path length $L$ for different values of the initial energy.  One can see that this effect is small ($\sim 2 \%$ ).  It does, however, have a positive correlation with the initial energy of the particle.

\subsection{The One Particle Analysis}

\begin{figure*}[t!]
\begin{center}
\hspace*{-.6in}
\psfig{file=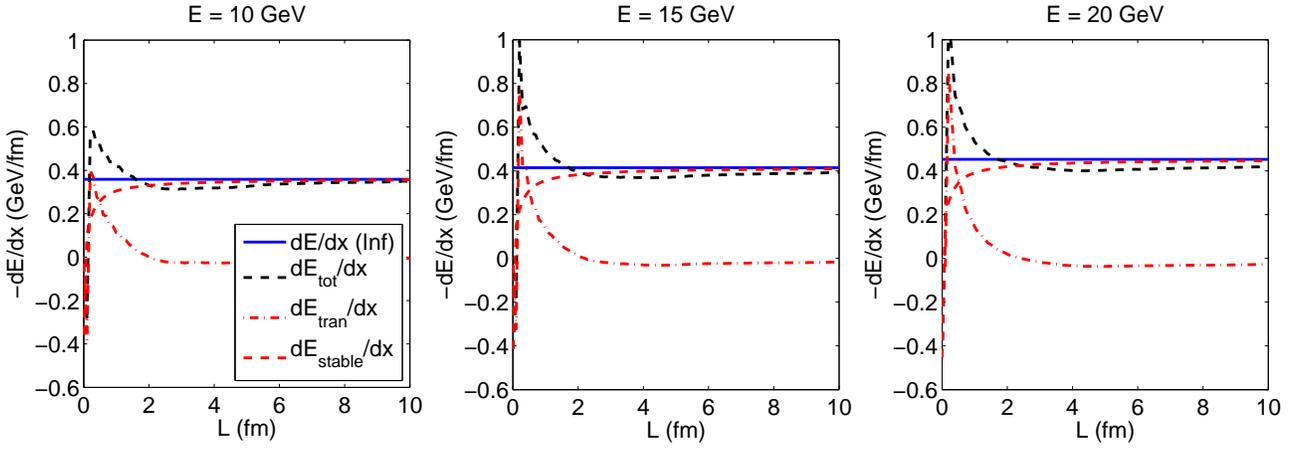,width=8in,angle=0}
\caption{(Color Online) The figures show the differential energy loss $-dE/dx$ suffered by a particle traveling in a medium as a function of the path length traversed $L$ for different values of the initial energy $E$.  The solid blue line is the infinite time calculation, the red dashed and dash-dotted curves are the stable and transient field energy losses respectively and the black dashed curve is the total finite time energy calculation.  Radiative modes are excluded from consideration.}
\label{fig:everythingdEdx}
\end{center}
\end{figure*}
\begin{figure*}[t!]
\begin{center}
\hspace*{-.6in}
\psfig{file=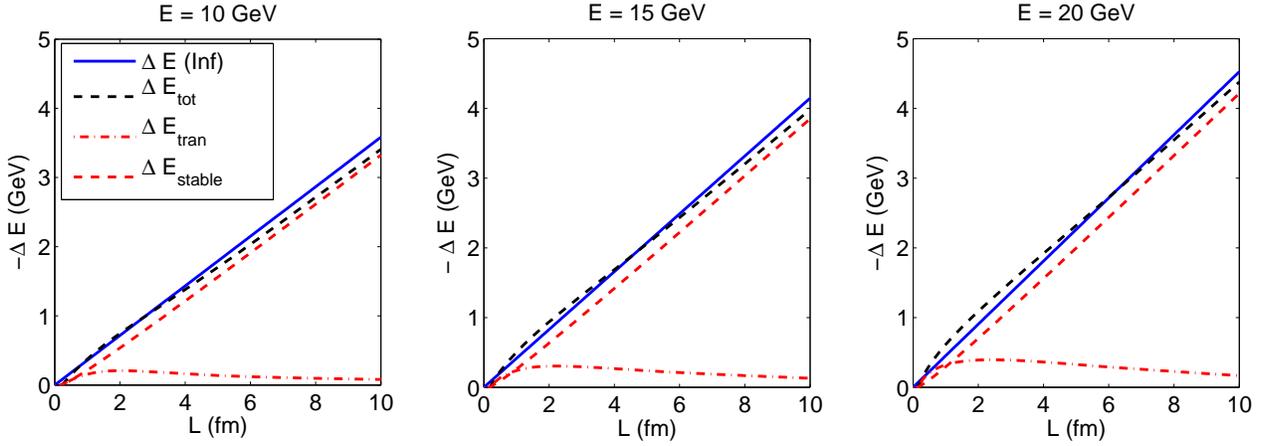,width=8in,angle=0}
\caption{(Color Online) The figures show the energy loss $-\Delta E$ suffered by a particle traveling in a medium as a function of the path length traversed $L$ for different values of the initial energy $E$.  The solid blue line is the infinite time calculation, the red dashed and dash-dotted curves are the stable and transient field energy losses respectively and the black dashed curve is the total finite time energy calculation.  Radiative modes are excluded from consideration.}
\label{fig:everythingdelE}
\end{center}
\end{figure*}
 The finite time energy loss can be schematically represented as $\Delta E \sim (D_{Ret}-D_{Rad})j$.  The removal of the radiative poles explicitly subtracts all radiative contributions.  We further write the retarded propagator in terms of the advanced and mixed parts, $\Delta E \sim (D_{Adv}+D_{-}-D_{Rad})j$.  
 As shown in Appendix \ref{App3}, the mixed propagator carries all the pole structure of the retarded propagator, only with twice the residue.  So we can further write $\Delta E \sim (D_{Adv}+A+2D_{Rad}-D_{Rad})j$.  The function $A$ is the cut contribution specified in Appendix \ref{App3}.  Using this procedure we can write the total finite time answer.
 \begin{eqnarray}
\frac{dE_{tot}}{dx}&=& \frac{dE_{tran}}{dx}+\frac{dE_{stable}}{dx} \nonumber \\
\frac{dE_{tran}}{dx}&=& \frac{1}{2}\left( \sum_{A\in \{pl\}}\frac{dE^{A}_{med}}{dx} - \sum_{B\in \{-\}}\frac{dE^{B}_{vac}}{dx} \right) \nonumber \\
\frac{dE_{stable}}{dx}&=&  \sum_{A\in \{cut,adv\}}\frac{dE^{A}_{med}}{dx}
\end{eqnarray}
We have broken up the one particle differential energy loss into a `transient' part and a `stable' part.  The `transient' field consists of the single set of simple poles that have been left over after the removal of the radiative energy loss.  The `stable' fields consist of the contributions coming from the particle-hole cut and the advanced part of the propagators.  Both these curves are shown in Figs.~\ref{fig:everythingdEdx} and \ref{fig:everythingdelE} along with the total energy loss and the infinite time calculation for comparison.  The curves are shown as a function of the path length $L$ and for different values of the initial energy $E$.

The `transient' fields are called as such because they are created at small initial times of the propagation in the medium and then quickly die away (as seen in Figs.~\ref{fig:everythingdEdx} and \ref{fig:everythingdelE}).  They tend to have a large effect since the current in use is created instantaneously out of the vacuum (with `infinite' acceleration).  Djordjevic \cite{djordj} finds similar contributions in the quantum mechanical calculation and attributes them to the lack of time translation invariance, another way of invoking initial conditions.

The `stable' field starts off at zero and then steadily builds up to the infinite case, reaching about $90 \%$ of the infinite field strength by a time on the order of one Debye length as one would expect (Fig.~\ref{fig:everythingdEdx}).  It is this delay in the set up of the fields by a time on the order of the Debye length which is the real retardation effect.  Fig. \ref{fig:everythingdelE} shows the integrated energy loss as a function of the path length traversed for different initial energies.  As one can see, the total energy loss in the finite case is of the same order as the infinite case and even slightly enhanced over the infinite case.  This enhancement comes from the contribution of the `transient' field effects that have not had time to settle down.  The `stable' field energy loss has a small depletion initially but builds up after a path length on the order of one Debye length and then the slope of the curve becomes equal to the infinite case.  This result is in agreement with the complementary calculation performed by Djordjevic \cite{djordj}.  A calculation with the current as formulated in Eq. \ref{physcurrent} would be useful in order to achieve an estimate for the magnitude of the transient effect.

\section{Conclusions}\label{Sec4}

We find that close attention must be paid to the form of the current used in the analysis using linear response formalism.  Some finite time currents lead to problems in interpretation and implementation of the formalism due to subtleties involved in ensuring that the current is conserved and that all radiative and binding effects have been discarded from the comparison being made to infinite time results.  The unphysically large binding effects are particularly important for the form of the current used by Peigne {\it et. al} \cite{peigne}, as a failure to account for them can cause a spurious interpretation of the results as a large time delay in the onset of collisional energy loss.

Our results show that there exists a small time delay (on the order of the Debye length) during which the chromoelectric fields build up to their full intensity.    Thus there is an initial depletion in energy loss in the finite time case as compared to the infinite case.  `Transient' field effects, however, push the energy loss further towards the infinite case; an effect that is sensitive to the detailed form of the current.  These `transient' fields, however, are unrealistically enhanced due to the form of the current and should be studied for more realistic charge configurations in order to estimate their true effects.  Our results are consistent with the calculations of Djordjevic \cite{djordj}, providing a complementary approach that both verifies and provides further physical insight into the result for non asymptotic heavy quark energy loss.

\begin{acknowledgments}
Discussions with M. Djordjevic, L. McLerran, B. Cole, 
I. Vitev and X.N. Wang
are gratefully acknowledged.  We are also grateful for correspondence with and clarifications provided by S. Peigne, P-B. Gossiaux and T. Gousset.
  This work is supported in part by the United States
Department of Energy
under Grants   No. DE-FG02-93ER40764.
\end{acknowledgments}

\appendix

\section{Solutions to Maxwell's Equations}\label{App1}

The solution to classical Yang-Mills theory in the Abelian approximation is equivalent to solving the Maxwell equations of electromagnetism.  In the vacuum the inhomogeneous part of the Maxwell equation can be written as the following.
\begin{equation}
\partial_{\mu}F^{\mu \nu a}=4\pi j^{\nu a}
\label{eq:maxwell}
\end{equation} 
Here, $F^{\mu \nu a}=\partial^{\mu}A^{\nu a}-\partial^{\nu}A^{\mu a}$ is the field strength tensor and $A^{\mu a}$ is the four vector potential (gluon field).  Note that the non linear part of the field strength has been omitted due to the Abelian approximation.  Equation \ref{eq:maxwell} can be solved in the Lorentz gauge and in fourier space to yield $A^{\mu a}(\omega,\vk)=-iD_{R}(\omega,\vk)j^{\mu a}(\omega,\vk)$ where we use the retarded Maxwell propagator, $D_{R}(\omega,\vk)=\frac{-4\pi i}{(\omega + i \eta)^{2}-k^{2}}$.  The electric field can be further calculated in Fourier space as the following.
\begin{equation}
\vE^{a}_{vac}(\omega,\vk)= D_{R}(\omega,\vk)(\omega \vj^{a} -\vk j^{0 a})
\label{eq:vacE}
\end{equation}

In the medium one has to solve the Maxwell equations with macroscopic fields in the linear response formalism.  We define the macroscopic fields ${\bf D}^{a}=\stackrel{\leftrightarrow}{{\bf \epsilon}} \cdot \vE^{a}$ and ${\bf B}^{a}=\stackrel{\leftrightarrow}{{\bf \mu}} \cdot {\bf H}^{a}$, with the permittivity and permeability tensors defined as the following: $\stackrel{\leftrightarrow}{{\bf \mu}}(\omega,\vk)=\mathbf{1}$ and $\stackrel{\leftrightarrow}{{\bf \epsilon}}(\omega,\vk)=\epsilon_{\perp}(\omega,\vk)(\mathbf{1}-\hat{{\bf k}}\hat{{\bf k}})+\epsilon_{\|}(\omega,\vk)\hat{{\bf k}}\hat{{\bf k}}$.  We solve the macroscopic Maxwell equations in Fourier space for the electric field.
\begin{eqnarray}
\vE^{a}(\omega,\vk)&=&\vE_{\perp}^{a}(\omega,\vk)+E_{\|}^{a}(\omega,\vk)\hat{\vk}\nonumber \\
\vE_{\perp}^{a}(\omega,\vk)&=&D_{R}^{\perp}(\omega,\vk)\vj_{\perp}^{a}(\omega,\vk) \nonumber \\
E_{\|}^{a}(\omega,\vk)&=&D_{\|}^{L}(\omega,\vk)j^{0 a}(\omega,\vk)
\label{eq:medE}
\end{eqnarray}
The transverse and longitudinal propagators are defined in terms of the transverse and longitudinal permittivities.
\begin{eqnarray}
D_{R}^{\perp}(\omega,\vk)&=&\frac{4\pi i \omega}{k^{2}-(\omega+i\eta)^{2}\epsilon_{\perp}(\omega+i\eta,\vk)}\nonumber \\
D_{R}^{L}(\omega,\vk)&=& \frac{-4\pi i}{k\epsilon_{\|}(\omega+i\eta,\vk)}
\label{eq:propmed}
\end{eqnarray}
The permittivities in turn can themselves be written in terms of the HTL one loop calculation for the polarization tensor, $\epsilon_{\perp}(\omega,\vk)=1-\Pi_{\perp}(\omega,\vk)/\omega^{2}$ and
$\epsilon_{\|}(\omega,\vk)=1+\Pi_{\|}(\omega,\vk)/k^{2}$.  The transverse and longitudinal polarizations are.
\begin{eqnarray}
\Pi_{\perp}(\omega,\vk)&=&\frac{1}{2}m_{D}^{2}\left(\frac{\omega}{k}\right)^{2}\left(1-\frac{\omega^{2}-k^{2}}{2\omega k}\log\left(\frac{\omega+k}{\omega-k}\right)\right) \nonumber \\
\Pi_{\|}(\omega,\vk)&=&m_{D}^{2}\left(1-\frac{\omega}{2k}\log\left(\frac{\omega+k}{\omega-k}\right)\right) 
\label{eq:pis}
\end{eqnarray}
Here $m_{D}^{2}=4\pi \alpha_{s}T^{2}(1+N_{F}/6)$ is the squared Debye (screening) mass where $T$ is the temperature of the medium and $N_{F}$ is the number of quark flavors that are in thermal equilibrium.

\section{Joule Heating and Radiative Energy Loss}\label{App2}

Joule heating is defined as the work done on a current by the electric field at its local position.  The total work done by the electric field on the current can be written as the following.
\begin{eqnarray}
W &=& \int dtd\vx  \vj^{a}(\vx,t)\cdot \vE ^{a}(\vx,t) \nonumber \\
&=&\int \invvol \vj^{a*}(\omega,\vk)\cdot\vE^{a}(\omega,\vk)
\label{eq:joule}
\end{eqnarray}
This expression can be used to calculate the work done by an electric field on its own source.  

Specifically for the vacuum case we can use Eq. \ref{eq:vacE} to write the following.
\begin{equation}
W= \int \invvol  D_{R}(\omega,\vk)(\omega(|\vj_{\perp}|^{2}+|j_{\|}|^{2})-k j_{\|}^{*}j^{0})
\label{eq:vacjoule}
\end{equation}
Note that all expressions quadratic in the current involve an implicit sum over the color index.  We separate the retarded propagator into a pole residue and a principal value part in order to isolate the radiative parts of the field.
\begin{eqnarray}
D_{R}(\omega,\vk)&=&PV \left (\frac{-4\pi i}{\omega^{2}-k^{2}} \right )+D_{rad}(\omega,\vk)\nonumber \\
D_{rad}(\omega,\vk)&=&-\frac{4\pi^{2}}{2k}(\delta(\omega-k)-\delta(\omega+k))
\label{eq:retpole}
\end{eqnarray}
We can isolate the radiative pole in the propagator in order to calculate the work done on the current due to the radiative field.
\begin{eqnarray}
W_{\textrm{rad}}&=&-\int \frac{d\vk}{4\pi^{2}}\{|\vj_{\perp}|^{2}+|j_{\|}|^{2}-\frac{1}{2}(j^{0*}j_{\|}+j_{\|}^{*}j^{0})\} \nonumber \\
&=&-\int \frac{d\vk}{4\pi^{2}}|\vj_{\perp}|^{2}
\label{eq:vacraden}
\end{eqnarray}
Eq. \ref{eq:vacraden} shows that the energy lost to Joule heating by the radiative part of the fields is exactly equal to the radiative energy loss of the current.  Note that we need a conserved current in order to get rid of the longitudinal current parts in Eq. \ref{eq:vacraden}.

\begin{figure}
\centering
 \epsfig{file=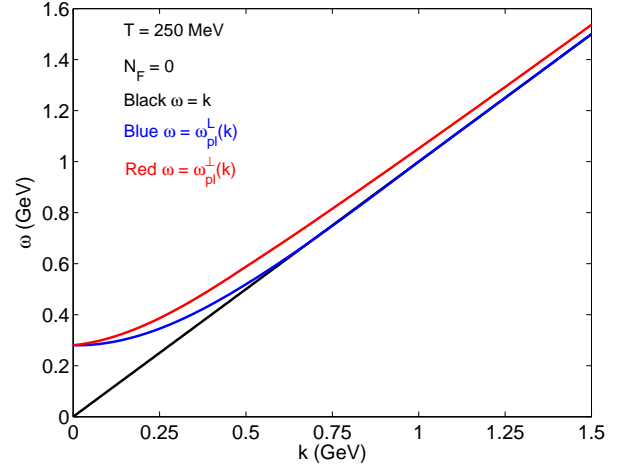,width=3.5in,angle=0}
  \caption{(Color Online) The figure shows the dispersion relations for the radiative fields in the vacuum and medium.  The calculation is done for $T=250$ MeV, $N_{F}=0$ and $\alpha_{s}=0.3$.  Note that $\omega_{pl}^{\perp}(0)=\omega_{pl}^{L}(0)=\frac{m_{D}}{\sqrt{3}}$}
 \label{fig:plasmon}
\end{figure}

We can do a similar calculation for the medium case in which we have separate longitudinal and transverse contributions to the radiation.  Using Eqs. \ref{eq:joule} and \ref{eq:medE} we can derive the following.
\begin{eqnarray}
W^{\perp}&=& \int \invvol D_{R}^{\perp}(\omega,\vk)|\vj_{\perp}|^{2} \nonumber \\
W^{\|}&=& \int \invvol D_{R}^{\|}(\omega,\vk)j_{\|}^{*}j^{0}
\label{eq:medjoule}
\end{eqnarray}
The medium retarded propagators can be separated into a radiative part and a principal value part in a way analogous to Eq. \ref{eq:retpole}.
\begin{eqnarray}
D_{R}^{\perp}(\omega,\vk)&=&PV(D_{R}^{\perp}(\omega,\vk))+D_{rad}^{\perp}(\omega,\vk) \nonumber \\
D_{rad}^{\perp}&=&-\frac{4\pi^{2}\omega}{g(\omega,\vk)}\left( \delta(\omega-\omega_{pl}^{\perp})+\delta(\omega+\omega_{pl}^{\perp})\right) \nonumber \\
D_{R}^{\|}(\omega,\vk)&=&PV(D_{R}^{\|}(\omega,\vk))+D_{rad}^{\|}(\omega,\vk) \nonumber \\
D_{rad}^{\|}&=&-\frac{4\pi^{2}k}{f(\omega,\vk)}\left( \delta(\omega-\omega_{pl}^{\|})+\delta(\omega+\omega_{pl}^{\|})\right)
\label{eq:radmedprop}
\end{eqnarray}
Here the functions $\omega_{pl}^{\perp}(k)$ and $\omega_{pl}^{\|}(k)$ are just the transverse and longitudinal plasmon dispersion relations respectively (shown in Fig. \ref{fig:plasmon}).  Note that the value of the plasmon frequency is always greater than the wave number ($\omega_{pl}^{L,\perp}\geq k$).  The functions $g(\omega,\vk)$ and $f(\omega,\vk)$ are defined as derivatives with respect to $\omega$ of the inverse propagators.
\begin{eqnarray}
g(\omega,\vk)&=&\partial_{\omega}(\omega^{2}-k^{2}-\Pi_{\perp}(\frac{\omega}{k})) \nonumber \\
&=& 2\omega-\frac{m_{D}^{2}\omega^{2}}{2k^{2}}\left(\frac{3}{\omega}-\frac{1}{k}\log\left(\frac{\omega+k}{\omega-k}\right)\right) \nonumber \\
f(\omega,\vk)&=&\partial_{\omega}(k^{2}+\Pi_{\|}(\frac{\omega}{k}))\nonumber \\
&=& m_{D}^{2}\left( \frac{\omega}{\omega^{2}-k^{2}}-\frac{1}{2k}\log\left(\frac{\omega+k}{\omega-k}\right)\right)
\end{eqnarray}

One can now determine how much work is done in joule heating due to fields of radiative type in the medium.  We can show using Eq. \ref{eq:radmedprop} inside Eq. \ref{eq:medjoule} that this amount is exactly equal to the zeroth order energy loss to plasmon radiation in the medium.
\begin{eqnarray}
W_{rad}^{\perp}&=&-\int \frac{d\vk}{4\pi^{2}}\frac{2\omega_{pl}^{\perp}}{g(\omega_{pl}^{\perp},\vk)}|\vj_{\perp}(\omega_{pl}^{\perp},\vk)|^{2}\nonumber \\
W_{rad}^{\|}&=&-\int \frac{d\vk}{4\pi^{2}}\frac{2k^{2}}{\omega_{pl}^{\|} f(\omega_{pl}^{\|},\vk)}|j_{\|}(\omega_{pl}^{\|},\vk)|^{2}
\label{eq:medraden}
\end{eqnarray}
We have used the continuity equation in fourier space to represent the longitudinal radiative energy loss purely in terms of the longitudinal current.  The expressions in Eq. \ref{eq:medraden} include the wave function renormalization factors due to interactions in the medium as well as the dependence on the current.

\section{Retarded, Advanced and Mixed Propagators}\label{App3}

In order to maintain causal effects, all fields in the paper are calculated from the sources using retarded boundary conditions.  Thus a field can be written schematically as $E=D_{R}j$.    We use a particular decomposition in order to separate the singularity structure of the propagator into something more amenable to physical interpretation.  Lets first illustrate this in the vacuum case.  We can write the retarded propagator in the following manner.
\begin{eqnarray}
D_{R}(\omega,\vk)&=&D_{R}(\omega,\vk)-D_{A}(\omega,\vk)+D_{A}(\omega,\vk)\nonumber \\
&=& D_{-}(\omega,\vk)+D_{A}(\omega,\vk)
\end{eqnarray}
In the previous expression we introduce the mixed propagator (which is just twice the half retarded half advanced Feynman propagator) $D_{-}(\omega,\vk)$.  The mixed propagator can be written as the following.
\begin{eqnarray}
D_{-}(\omega,\vk)&=&D_{R}(\omega,\vk)-D_{A}(\omega,\vk)\nonumber \\
&=& -\frac{4\pi^{2}}{k}\left( \delta(\omega-k)-\delta(\omega+k)\right)
\end{eqnarray}
Note that the advanced and retarded vacuum propagators are only different when it comes to the position of singularities in the complex plane.  Taking their difference, as the mixed propagator does, just picks out this singularity structure.  Note that this contribution is just twice the radiative contribution.  We will see that this same procedure with the medium propagators will lead to the isolation of the particle hole cut as well as the plasmon pole.

The basic difference between the vacuum and medium cases is the existence of the branch cut singularity in the medium propagators that is not present in the vacuum case.  The principal value part that simply cancels out in the vacuum mixed propagator actually produces this electron-hole pair cut in the medium.
\begin{eqnarray}
D_{-}^{\perp}(\omega,\vk)&=&A^{\perp}(\omega,\vk)+2D_{rad}^{\perp}(\omega,\vk) \nonumber \\
A^{\perp}(\omega,\vk)&=&\frac{2\pi^{2}m_{D}^{2}\omega^{2}(\omega^{2}-k^{2})\Theta(k+\omega)\Theta(k-\omega)}{k^{3}(k^{2}-\omega^{2}+\Pi_{\perp}(\frac{\omega+i\eta}{k}))(k^{2}-\omega^{2}+\Pi_{\perp}(\frac{\omega-i\eta}{k}))} \nonumber \\
D_{-}^{\|}(\omega,\vk)&=&A^{\|}(\omega,\vk)+2D_{rad}^{\|}(\omega,\vk) \nonumber \\
A^{\|}(\omega,\vk)&=&\frac{-4\pi^{2}m_{D}^{2}\omega\Theta(k+\omega)\Theta(k-\omega)}{(k^{2}+\Pi_{\|}(\frac{\omega+i\eta}{k}))(k^{2}+\Pi_{\|}(\frac{\omega-i\eta}{k}))}
\end{eqnarray}
The functions $A^{\perp}(\omega,\vk)$ and $A^{\|}(\omega,\vk)$ occur as a result of the branch cut singularities present in the medium propagators.  They can be interpreted as a contribution from the particle-hole continuum present in the plasma.  Note the difference between these branch cuts and the cuts in the propagators used in Thoma-Gyulassy \cite{tg} as only half the cut is used in that paper.

%\begin{thebibliography}{80}

%\end{thebibliography}

\end{document}